\documentclass[10pt,conference]{IEEEtran}
\usepackage{cite}
\usepackage{amsmath,amssymb,amsfonts}
\usepackage{graphicx}
\usepackage{textcomp}
\usepackage{xcolor}
\usepackage[utf8]{inputenc}
\usepackage{hyperref}
\usepackage{booktabs}
\usepackage[switch]{lineno}
\definecolor{softgreen}{RGB}{102, 194, 165} 
\definecolor{softred}{RGB}{252, 141, 98}   

\def\BibTeX{{\rm B\kern-.05em{\sc i\kern-.025em b}\kern-.08em
    T\kern-.1667em\lower.7ex\hbox{E}\kern-.125emX}}

\usepackage{xcolor}

\usepackage{multirow} 

\begin{document}
%
%
\title{Disguising Topology and Side-Channel Information through Covert Gate- and ML-Enabled IP Camouflaging}
\author{\IEEEauthorblockN{Junling Fan, David Koblah, Domenic Forte}
\IEEEauthorblockA{\textit{Department of Electrical and Computer Engineering} \\
\textit{University of Florida}, Gainesville, USA \\
{\{fan.j, dkoblah\}@ufl.edu, dforte@ece.ufl.edu}
}}
%
%
%


\maketitle


\begin{abstract}
Semiconductor intellectual property (IP) theft incurs hundreds of billions in annual losses, driven by advanced reverse engineering (RE) techniques. Traditional ``cryptic'' IC camouflaging methods typically focus on hiding localized gate functionality but remain vulnerable to system-level structural analysis. This paper explores ``mimetic deception,'' where a functional IP ($F$) is designed to structurally and visually masquerade as a completely different appearance IP ($A$). We provide a comprehensive evaluation of three deceptive methodologies: IP Camouflage, Graph Matching, and DNAS-NAND Gate Array, analyzing their resilience against GNN-based node classification, and Differential Power Analysis (DPA). Crucially, we demonstrate that mimetic deception achieves a novel anti-side-channel defense: by forcing the mis-classification of cryptographic primitives, the adversary is led to apply an incorrect power model, causing the DPA attack to fail. Our results validate that this multi-layered approach effectively thwarts the entire RE toolchain by poisoning the structural and logical data used for netlist understanding.
\end{abstract}
\renewcommand\IEEEkeywordsname{Keywords}

%

\section{Introduction}

The integrity of modern electronic hardware is increasingly threatened by physically invasive reverse engineering (RE). By systematically delayering and imaging integrated circuits (ICs), attackers can reconstruct gate-level netlists to extract sensitive intellectual property (IP), identify security vulnerabilities, or facilitate illegal cloning. This poses catastrophic risks to critical domains, including defense and finance, where the exposure of underlying algorithms can undermine national and economic security~\cite{RE2016,arancaGlobalResearch}.

IC camouflaging has emerged as a key defense, utilizing fabrication-level secrets, such as dummy contacts~\cite{rajendran2013security}, threshold voltage~\cite{akkaya2018secure} or technology~\cite{shakya2019covert}, to make different logic gates appear identical under Scanning Electron Microscope (SEM) imaging. However, existing methods often focus on localized gate modifications and fail to achieve broader system-level deception. This narrow focus allows advanced AI-enhanced RE tools and SAT solvers to eventually de-obfuscate the design.

This paper evaluates a holistic methodology known as mimetic deception~\cite{fan2025designing,fan2025scalable}, which integrates cryptic and mimetic principles to ensure a circuit not only hides its true function but also presents a realistic yet misleading outward appearance. We evaluate three primary methods:
\begin{itemize}
    \item \textbf{IP Camouflage:} An ML-driven approach utilizing an And-Inverter Graph Variational Autoencoder (AIG-VAE) to blend the functionality of a circuit with a deceptive appearance.
    \item \textbf{Graph Matching:} A layer-by-layer greedy heuristic that operates on standard logic gates to map a functional netlist onto a decoy topology while maintaining formal equivalence
    \item \textbf{DNAS-NAND Gate Array:} A scalable synthesis model using Differentiable Neural Architecture Search (DNAS) to generate end-to-end deceptive circuits within a single-stage flow.
\end{itemize}

Beyond traditional metrics of SAT resilience and PPA overhead, this work focuses on a unique defensive vector: side-channel misclassification. We analyze how mimetic deception can disguise cryptographic primitives to cause the failure of Differential Power Analysis (DPA). By making one cipher (e.g., PRESENT) appear as another (e.g., AES or DES), the attacker is misled into using an incompatible hypothetical power model. Since the success of DPA relies on the precise correlation between the power trace and the specific logic of the algorithm, this functional misclassification prevents the adversary from ever reaching the true secret key.

\section{Background}
\begin{figure*}[ht]
    \centering
    \includegraphics[width=\linewidth]{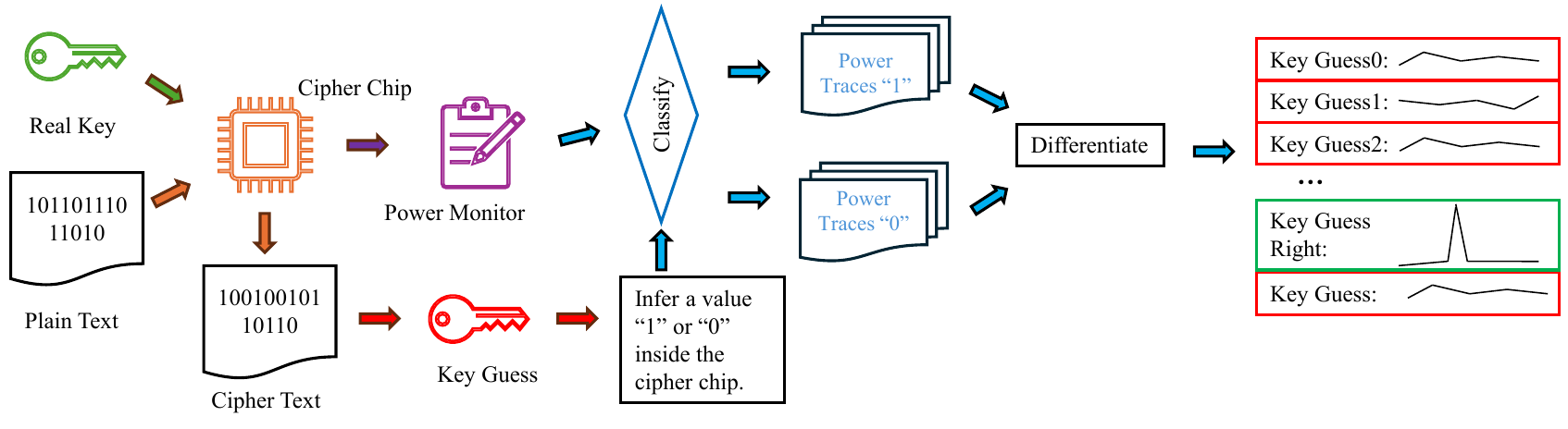}
    
    \caption{Overview of a Differential Power Analysis (DPA) attack workflow targeting a cryptographic S-Box with a peak for correct key guess.}
    \label{fig:dpa_workflow}
\end{figure*}
\subsection{IC Camouflaging and Covert Gates}
IC camouflaging techniques traditionally fall into two categories: gate-level and interconnect-level~\cite{shakya2019covert}. Gate-level methods replace standard cells with variants that implement different functions based on fabrication secrets like dummy contacts~\cite{rajendran2013security} or threshold voltage variations~\cite{erbagci2016secure}. Interconnect camouflaging~\cite{yu2017incremental} obscures signal paths by manipulating internal wiring.

The current state-of-the-art is the covert gate methodology~\cite{shakya2019covert}. Covert gates exploit always-on and always-off transistors to realize multiple gate functionalities within a single physical footprint. Because these transistor states are created through doping and remain indistinguishable under SEM imaging, covert gates appear identical to standard CMOS cells. This work utilizes specialized variants: Fake Inverters (FI), Fake Buffers (FB), and Universal Transmitters (UT), implementing function-appearance mismatches that thwart both visual inspection and SAT-based logical attacks~\cite{fan2025designing}.
\subsection{IC Reverse Engineering and Netlist Understanding}
Physically invasive reverse engineering (RE) is a destructive process where an attacker systematically dissects an IC to uncover its complete netlist~\cite{torrance2011state,quadir2016survey}. The typical workflow begins with decapsulation to expose the silicon die, followed by iterative deprocessing. During this phase, each metal and dielectric layer is removed through mechanical polishing or automated ion beam milling~\cite{principe2017steps}. High-resolution SEM imaging is then used to capture the layout, which is subsequently analyzed to extract a gate-level netlist~\cite{quadir2016survey}.

Once extracted, netlist understanding tools are employed to abstract high-level functions from the flattened ``sea of gates"~\cite{albartus2020dana}. Algorithmic approaches identify recurring structures or infer control logic through Boolean reasoning~\cite{subramanyan2013reverse, meade2016netlist}. More recently, machine learning-based tools such as GNN-RE~\cite{alrahis2021gnn} and FGNN2~\cite{wang2024fgnn2} interpret circuits as graphs to classify gate functions and functional boundaries. Mimetic deception aims to poison the structural data these tools rely on, forcing them to produce misleading architectural labels.

\subsection{DPA Attack on Ciphers}
Differential Power Analysis (DPA) is a statistical side-channel attack that exploits the correlation between a device's instantaneous power consumption and the intermediate data values it processes~\cite{kocher1999differential}. Unlike Simple Power Analysis (SPA), which relies on visual inspection, DPA utilizes large datasets to extract secret keys even amidst significant noise~\cite{prouff2005dpa}.

The attack is predicated on a power leakage model, typically the Hamming Weight (HW) or Hamming Distance (HD) model. As detailed by Prouff~\cite{prouff2005dpa}, the power consumption $C(X)$ of a device processing data $X$ is modeled as proportional to the number of bit transitions (HD) or the number of set bits (HW). In block ciphers like AES or DES, the Substitution Box (S-Box) is the primary target due to its non-linearity, which relates plaintext, key, and ciphertext in a complex but deterministic manner.

To perform a DPA attack, an adversary defines a \textit{selection function} $D(X, K, j)$, which predicts the value of the $j$-th bit of a specific intermediate state (e.g., the S-Box output) given a plaintext $X$ and a hypothetical key $K$. The adversary then computes the \textit{differential trace} $\Delta_{K}$ by partitioning the measured power traces into two sets based on whether $D$ predicts a 0 or a 1, and calculating the difference of their means:
\begin{equation}
\begin{split}
    \Delta_{K}(j) &= \frac{\sum_{i=1}^{N} D(X_i, K, j) C(X_i)}{\sum_{i=1}^{N} D(X_i, K, j)} \\
    &\quad - \frac{\sum_{i=1}^{N} (1-D(X_i, K, j)) C(X_i)}{\sum_{i=1}^{N} (1-D(X_i, K, j))}
\end{split}
\label{eq:differential_trace}
\end{equation}
\noindent where $N$ is the number of traces and $C(X_i)$ is the measured power consumption for the $i$-th input.

When the hypothetical key $K$ matches the actual secret key, the selection function $D$ accurately classifies the power traces, resulting in a statistically significant peak in $\Delta_{K}$ at the time instant the intermediate value is processed -- see Fig.~\ref{fig:dpa_workflow}. Conversely, incorrect key hypotheses result in uncorrelated groupings, where the difference of means tends toward zero. However, as noted in~\cite{prouff2005dpa}, incorrect keys can sometimes produce ``ghost peaks" due to residual correlations, though these are typically distinguishable from the correct key signal with sufficient data.

\textit{Crucially, the success of this methodology relies entirely on the correctness of the selection function $D$. The attacker must precisely know which cryptographic primitive (e.g., AES S-Box vs. PRESENT S-Box) is executing to accurately predict the intermediate bit $j$. If the target logic is obfuscated such that the attacker employs a selection function derived from a decoy S-Box, the predictions will fail to correlate with the physical power leakage, rendering the attack futile.}
\section{Deceptive Design Methodologies}
\label{sec:methodologies}

To realize the paradigm of mimetic deception, we propose and evaluate three distinct methodologies. Each approach aims to synthesize a circuit that performs a target function ($F$) while mimicking the structural topology of a decoy appearance circuit ($A$). However, they differ fundamentally in their algorithmic foundations -- ranging from heuristic graph theory to generative machine learning -- and their respective trade-offs in scalability, overhead, and representation.

\begin{figure*}[t]
    \centering
    \includegraphics[width=\linewidth]{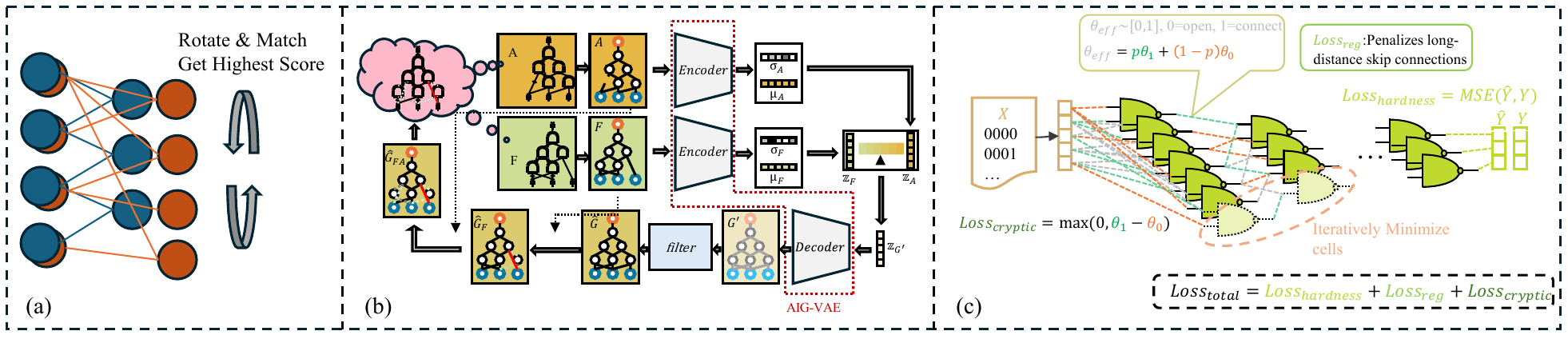}
    \caption{Overview of the three proposed deceptive design methodologies: (a) Graph Matching using standard gates, (b) IP Camouflage using AIG-VAE interpolation, and (c) DNAS-Based NAND Array optimization.}
    \label{fig:methods_overview}
\end{figure*}

\subsection{IP Camouflage}
IP Camouflage is the foundational generative approach for mimetic deception~\cite{fan2025designing}. Unlike heuristic matching, this method utilizes deep learning to capture and manipulate the latent functional representations of circuits.

The workflow employs a novel And-Inverter Graph Variational Autoencoder (AIG-VAE). The encoder utilizes an asynchronous message-passing mechanism and Gated Recurrent Units (GRUs) to encode the structural and functional dependencies of an AIG into a probabilistic latent space defined by a mean $\mu$ and variance $\sigma$~\cite{fan2025designing}. To generate a deceptive circuit, the model interpolates between the latent vector of the functional circuit ($\mathbb{Z}_F$) and the appearance circuit ($\mathbb{Z}_A$) using a proportion factor $p \in [0,1]$:
\begin{equation}
    \mathbb{Z}_{G'} = (1-p)\mathbb{Z}_F + p\mathbb{Z}_A
\end{equation}
The decoder then reconstructs a ``blended" circuit from $\mathbb{Z}_{G'}$ by predicting node types, connections, and inverter placements~\cite{fan2025designing}.

Because the generative output is a probabilistic approximation, this method requires a critical ``Post-Generation Rectification'' phase to restore logical integrity. In this step, the blended circuit is compared against the target function $F$. Discrepancies are systematically resolved by inserting specialized covert gates:
\begin{itemize}
    \item \textbf{Fake Inverters (FI) and Fake Buffers (FB):} Components that appear as standard inverters/buffers but function as constant logic or wires~\cite{fan2025designing}.
    \item \textbf{Universal Transmitters (UT):} A flexible component that appears as a NAND gate but can be configured to act as a buffer, inverter, or constant source~\cite{fan2025designing}.
\end{itemize}
While this approach effectively creates function-appearance mismatches, the reliance on post-generation fixes introduces area overhead and limits the method's scalability to smaller logic cones (approx. 200 nodes)~\cite{fan2025designing}.

\subsection{Graph Matching}
The Graph Matching method addresses the representation mismatch inherent in AIG-based approaches~\cite{fan2025designing} by operating directly on standard logic gate netlists~\cite{fan2025scalable}. Unlike AIGs, which decompose logic into abstract AND/NOT nodes, this method processes the ``real logic" representation found in standard Process Design Kits (PDKs). This ensures that the generated deceptive circuit is composed of standard cells (e.g., NAND, NOR, XOR) that are directly compatible with industrial design flows and standard reverse engineering tools.

The core of this method is a layer-by-layer greedy heuristic that maps nodes from the appearance graph $G_A$ to the functional graph $G_F$~\cite{fan2025scalable}. The algorithm proceeds in three distinct phases:

\begin{enumerate}
    \item \textbf{Graph Levelization:} Both the functional ($G_F$) and appearance ($G_A$) graphs are independently levelized using a topological sort (Kahn's algorithm). This assigns every node to a specific depth layer, establishing a structured basis for matching~\cite{fan2025scalable}.
    \item \textbf{Bipartite Matching:} The algorithm iterates through the circuit depth from inputs to outputs. At each layer $k$, it solves a minimum-cost bipartite matching problem to map nodes from $Layer_k(G_A)$ to $Layer_k(G_F)$ using the Hungarian Algorithm. This ensures a globally optimal assignment for that specific layer before propagating constraints to the next~\cite{fan2025scalable}.
    \item \textbf{Asymmetric Cost Optimization:} The matching decisions are driven by a specialized cost function designed to enforce deception. The cost $C_{i,j}$ of matching an appearance node $a_i$ to a functional node $f_j$ is the sum of two components:
    \begin{itemize}
        \item \textit{Node Cost:} A penalty is derived from the difficulty of camouflaging the logic function. For example, hiding a simple buffer or inverter within a complex NAND/NOR gate incurs zero cost, as supported by Covert Gate technology. However, matching incompatible logic types incurs a high penalty~\cite{fan2025scalable}.
        \item \textit{Connection Cost:} This enforces a structural containment constraint based on the matching from the previous layer ($M_{prev}$). A high penalty is assigned if a connection exists in the functional graph $G_F$ but the corresponding mapped nodes in $G_A$ are not connected. This effectively encourages the connectivity of $G_F$ to be a structural subset of $G_A$, allowing the functional wires to be "hidden" within the appearance topology~\cite{fan2025scalable}.
    \end{itemize}
\end{enumerate}

\subsection{DNAS-Based Gate Array Optimization}
To overcome the scalability limitations of the AIG-VAE and the representation constraints of standard AIGs, we introduced the DNAS-NAND Gate Array method~\cite{fan2025scalable}. This approach leverages Differentiable Neural Architecture Search (DNAS) to synthesize an end-to-end deceptive circuit without the need for post-generation rectification.

We proposed a unified model named ``SelectorTNet'', which extends the T-Net~\cite{wang2024towards} architecture to learn two distinct logic functions simultaneously. The network incorporates a global selector variable $p$ that governs the effective parameters $\theta_{eff}$ of the network during the forward pass:
\begin{equation}
    \theta_{eff} = p \cdot \theta_{Functional} + (1-p) \cdot \theta_{Appearance}
\end{equation}
When $p=1$, the network parameters adapt to synthesize the functional circuit; when $p=0$, they adapt to the appearance circuit. The use of Gumbel-Softmax sampling ensures the selection of connections remains differentiable~\cite{fan2025scalable}.

The training is guided by a composite loss function that jointly optimizes for correctness and mimicry:
\begin{itemize}

    \item \textbf{Hardness-Aware Loss:} Ensures 100\% logical correctness for both $F$ (at $p=1$) and $A$ (at $p=0$) by penalizing ``hard" examples where the network's prediction diverges from the truth table~\cite{fan2025scalable}.
    \item \textbf{Inner-Architecture Regularized Loss:} Penalizes long-distance skip connections to ensure the synthesized netlist is physically realizable and efficient~\cite{fan2025scalable}.
    \item \textbf{Cryptic Loss (Connection Containment):} This is the core mechanism for deception. It penalizes any connection in the functional mode ($p=1$) that exhibits a stronger activation magnitude than its counterpart in the appearance mode ($p=0$). The loss is defined as:
    \begin{equation}
        Loss_{cryptic} = \max(0, \theta_{1} - \theta_{0})
    \end{equation}
\end{itemize}
This constraint effectively forces the functional circuit to be ``hidden" within the wiring of the appearance circuit. The result is a single NAND-gate array that satisfies the functional requirements while structurally appearing as the decoy, achieving high scalability and ultra-low overhead~\cite{fan2025scalable}.

\subsection{Summary of Methodologies}
\label{subsec:MethodSum}
Table~\ref{tab:method_comparison} summarizes the key characteristics of the three proposed methods. Graph Matching excels in formal correctness and standard representation, but offers moderate overhead due to strict matching constraints. IP Camouflage pioneered the generative approach but is limited by the rectification bottleneck. The DNAS-NAND Array offers the best scalability and PPA efficiency, albeit with a homogeneous NAND-only representation and difficulty in accurate representation of all functions such as ciphers.

\begin{table}[t]
\scriptsize
\caption{Comparison of Deceptive Design Methodologies}
\begin{center}
\begin{tabular}{lccc}
\toprule
\textbf{Feature} & \textbf{Graph Matching} & \textbf{IP Camouflage} & \textbf{DNAS-NAND} \\
\midrule
\textbf{Core Algo.} & Greedy Heuristic & AIG-VAE & DNAS \\
\textbf{Logic Basis} & Standard Gates & AIGs & NAND Array \\
\textbf{Scalability} & High & Low ($<200$ nodes) & High \\
\textbf{Rectification} & Not Required & Required & Not Required \\
\textbf{Formal Eq.} & Guaranteed & Guaranteed & Verified ex-post \\
\textbf{Target} & Arithmetic/Cipher & Arithmetic/Cipher & Arithmetic \\
\bottomrule
\end{tabular}
\label{tab:method_comparison}
\end{center}
\end{table}

In this work, we target the S-Boxes of varying Ciphers, distinct for their high non-linearity. Although test-vector based synthesis~\cite{belcak2022neural, wang2024towards, fan2025scalable} is efficient for arithmetic circuits, it struggles to encompass the complex behavior of S-Boxes without incurring unacceptable hardware costs. As demonstrated in~\cite{telgarsky2016benefits}, the number of neurons required to approximate a function grows exponentially with its number of oscillations. For highly non-linear S-Boxes, this theoretical bound manifests as a massive area overhead, rendering standard learning-based synthesis ineffective.
\section{Experiments}
\label{sec:experiments}

\begin{table*}[t]
\centering
\caption{Comparison of Deception Metrics and Overhead: IP Camouflage (AIG) vs. Graph Matching (GM)}
\label{tab:results}
\resizebox{\textwidth}{!}{%
\begin{tabular}{|c|c|c|c|c|c|c|c|c|}
\hline
\multirow{2}{*}{\textbf{Scenario ($F \rightarrow A$)}} & \multirow{2}{*}{\textbf{Method}} & \multicolumn{2}{c|}{\textbf{Deception Scores}} & \multicolumn{2}{c|}{\textbf{GNN F1 Scores}} & \textbf{DPA Key Hiding} & \multicolumn{2}{c|}{\textbf{Overhead (Normalized)}} \\ \cline{3-9} 
 &  & \textit{$Score_{GNNRE}$} & \textit{$Score_{DPA}$} & \textit{$F1_{expose}$} & \textit{$F1_{mimicry}$} & \textit{$Rank_{Hidden}$} & \textit{Area} & \textit{Power} \\ \hline

\multirow{2}{*}{$PRESENT \rightarrow DES$ (Avg)} & AIG & 3.32 & 0.47 & 0.10 & 0.37 & 34.1/64 & 1.12 & 1.11 \\ \cline{2-9} 
 & GM & 9.40 & 0.44 & 0.05 & 0.32 & 33.6/64 & 1.14 & 1.25 \\ \hline

\multirow{2}{*}{$PRESENT \rightarrow AES$} & AIG & 18.0 & 0.43 & 0.05 & 0.95 & 115.2/256 & 1.02 & 1.02 \\ \cline{2-9} 
 & GM & 7.40 & 0.44 & 0.10 & 0.84 & 120.2/256 & 1.05 & 1.17 \\ \hline

\multirow{2}{*}{$DES \rightarrow AES$ (Avg)} & AIG & $\infty$ (High) & 0.45 & 0.005 & 0.90 & 136.1/256 & 1.07 & 1.07 \\ \cline{2-9} 
 & GM & 84.6 & 0.38 & 0.015 & 0.89 & 124.6/256 & 1.11 & 1.12 \\ \hline
\end{tabular}%
}
\vspace{2mm}
\footnotesize{\\ \textbf{Note:} $F \rightarrow A$ denotes Functional Circuit disguised as Appearance Circuit. 'Avg' represents the mean across all 8 S-Box variants ($S1-S8$). Score\_GNNRE '$\infty$' indicates the classifier failed to converge or produced random guess equivalent results. In best scenario, $Rank_{Hidden} = 0.5$}.
\end{table*}

\begin{figure}[t]
    \centering
    \includegraphics[width=0.99\linewidth]{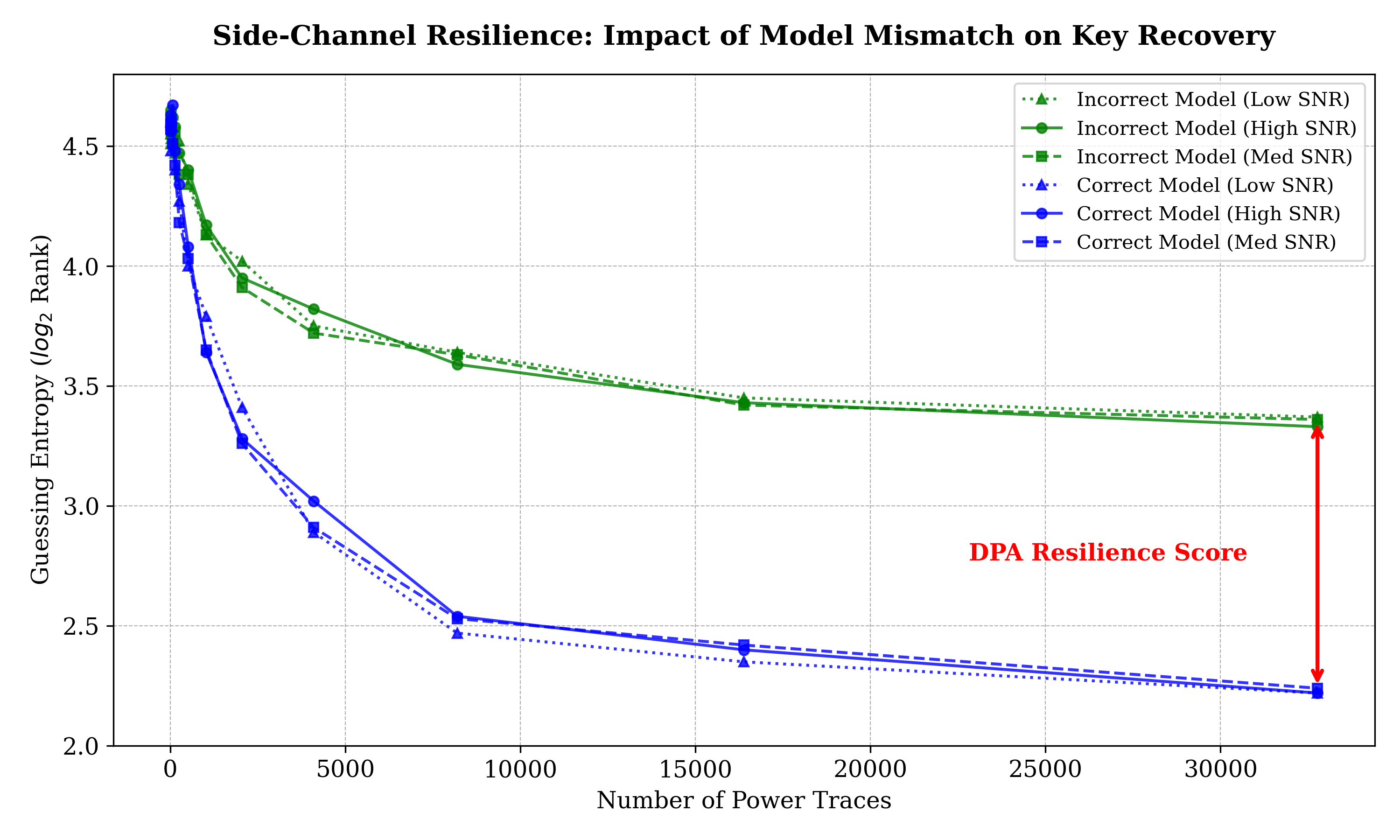}
    \caption{Evaluation of DPA Resilience under Mimetic Deception. The plot compares the Guessing Entropy (GE) of a target S-Box against an attacker using the correct power model (Blue/Baseline) versus an incorrect model induced by deception (Green/Deceptive). While the baseline attack rapidly converges to the correct key (low entropy) within 2,000 traces, the deceptive design maintains high entropy even after 32,000 traces, preventing successful key recovery. The red arrow indicates the DPA Resilience Score, quantifying the sustained protective gap achieved by forcing a statistical model mismatch.}
    \label{fig:dpa_example}
\end{figure}

We evaluate the efficacy of the proposed mimetic deception methodologies, IP Camouflage (AIG) and Graph Matching (GM), against two primary threats: automated reverse engineering via Graph Neural Networks (GNNs) and recovering secret key by Differential Power Analysis (DPA).

\subsection{Experiment Setup \& Scope}
Our experiments target the highly non-linear Substitution Boxes (S-Boxes) of cryptographic primitives, specifically PRESENT (4-bit), DES (6-bit to 4-bit), and AES (8-bit).

\vspace{0.5ex}

\noindent \textbf{Synthesis and Attack:} All designs were camouflaged, modeled and synthesized using SAED 90nm Cell Library. GNN-based structural analysis was performed using a state-of-the-art GNN-RE classifier~\cite{alrahis2021gnn}. Side-channel resilience was evaluated using a Differential Power Analysis (DPA) attack framework.

\vspace{0.5ex}

\noindent \textbf{Exclusion of DNAS-NAND Method:} While~\autoref{sec:methodologies} introduced the DNAS-NAND Gate Array as a scalable candidate for arithmetic circuits, we exclude it from this experimental evaluation of S-Boxes. As theoretically argued in~\autoref{subsec:MethodSum}, emulating highly non-linear functions (like cryptographic S-Boxes) with a shallow neural-based synthesis model requires a prohibitively large number of neurons to capture the high-frequency oscillations. Preliminary trials confirmed that trying to synthesize these S-Boxes via DNAS resulted in area overhead exceeding $10\times$, rendering the method impractical for cryptographic primitives compared to AIG and GM.

\subsection{Side-Channel Resilience Analysis}

To quantify the mimetic deception effect on side-channel security, we disguised a PRESENT S-Box to structurally appear as a DES S-Box. We then performed a DPA attack assuming the adversary relies on the visual appearance to select the power model. \autoref{fig:dpa_example} plots the Guessing Entropy (GE) defined as the expected rank of the secret key against the number of power traces (up to 32k).

\begin{itemize}
    \item \textbf{Baseline (Correct Model):}  When the attacker uses the correct power model (Blue lines), the attack succeeds rapidly. The guessing entropy drops logarithmically, approaching $\approx 2.2$ bits (rank $<5$) with 32k traces, indicating the key is effectively recovered.
    \item \textbf{Deceptive (Incorrect Model):} When the attacker is misled into using the model of the appearance circuit (Green lines), the attack fails to converge. Even with 32k traces, the guessing entropy remains high ($\approx 3.4$ bits), meaning the correct key remains hidden among a large pool of candidates.
\end{itemize}

\subsection{Comparative Analysis: AIG vs. GM}
\label{subsec:CompAnalysis}

In this section, we evaluate the performance of the proposed methods using two novel security metrics designed to quantify the effectiveness of mimetic deception in both the side-channel and structural domains.

\subsubsection{Side-Channel Resilience (DPA Score)}
To measure the ``protective gap'' created by the function-appearance mismatch as shown in~\autoref{fig:dpa_example}, we define the \textbf{DPA Resilience Score} as the normalized divergence in key ranking between the correct and incorrect power models. It is calculated as:
\begin{equation}
    Score_{DPA} = \frac{Rank_{disguise} - Rank_{leak}}{Rank_{max}}
    \label{eq:dpa_score}
\end{equation}
where $Rank_{leak}$ is the final rank of the secret key under the correct model (baseline), $Rank_{disguise}$ is the rank under the deceptive model, and $Rank_{max}$ is the size of the key hypothesis space (e.g., 64 for DES, 256 for AES).

Ideally, this score approaches \textbf{0.5}. This theoretical maximum is achieved when the baseline attack succeeds perfectly ($Rank_{leak} \approx 0$) while the deceptive attack forces the adversary into a state of random guessing, where the key rank settles at the statistical median ($Rank_{disguise} \approx 0.5 \cdot Rank_{max}$). A score of 0.5 indicates the key is hidden in a ``middle ground,'' indistinguishable from noise.

As shown in Table~\ref{tab:results}, both methods achieve scores consistently in the range of \textbf{0.38 -- 0.47}. This confirms that regardless of the synthesis method (Generative AIG or Heuristic GM), the topological deception successfully masks the leakage, keeping the key rank high and preventing convergence.

\subsubsection{Structural Deception (GNN Score)}
To quantify resilience against automated structural analysis, we utilize the \textbf{GNN Deception Score}~\cite{fan2025scalable} as defined in Eq.~(3):
\begin{equation}
    Score_{GNN} = \frac{F1_{mimicry} - F1_{expose}}{F1_{expose}}
    \label{eq:gnn_score}
\end{equation}
where $F1_{expose}$ measures the accuracy of the GNN in identifying the hidden functional logic, and $F1_{mimicry}$ measures its success in misclassifying the logic as the decoy appearance. A higher score indicates that the tool is significantly more likely to be misled by the outer topology ($A$) than to uncover the inner functionality ($F$).

\begin{itemize}
    \item \textbf{IP Camouflage (AIG)} demonstrated peak performance, achieving ``Infinite'' scores in several DES$\to$AES test cases. This occurs as $F1_{expose} \to 0$, representing total obfuscation where the GNN fails to identify the functional logic entirely.
    \item \textbf{Graph Matching (GM)} yielded consistent high scores (e.g., $\approx 84.6$ for DES$\to$AES), proving that standard-gate mapping effectively poisons the structural features relied upon by RE classifiers.
\end{itemize}

\subsubsection{Overhead and Fidelity Trade-offs}
We evaluate PPA metrics normalized against the \textit{Appearance Circuit} ($A$) to quantify the deviation from the target topology. A value of $1.0\times$ represents ideal mimicry where the deceptive circuit is physically indistinguishable from the decoy.
\begin{itemize}
    \item \textbf{IP Camouflage (AIG)} demonstrates superior density, maintaining an area profile near-identical to the target ($1.02\times - 1.12\times$). This efficiency stems from the AIG-VAE's latent-space interpolation, which naturally generates compact structures.
    \item \textbf{Graph Matching (GM)} incurs slightly higher deviations ($1.14\times$ Area, $1.25\times$ Power) due to the strict bipartite matching constraints. To guarantee formal equivalence using standard gates, GM often requires the insertion of additional dummy chains (covert gates~\cite{shakya2019covert}) to satisfy the connectivity requirements of the appearance graph, resulting in a slight bloating of the power footprint.
\end{itemize}

\section{Conclusion}
\label{sec:conclusion}

This work presents a comprehensive evaluation of ``mimetic deception,'' a defensive paradigm that decouples a circuit's logical function from its physical topology to thwart both reverse engineering and side-channel analysis. By disguising the S-Boxes of vulnerable ciphers (e.g., PRESENT, DES) to structurally mimic robust or unrelated primitives (e.g., AES), we successfully mislead the automated toolchains used by adversaries.

Our experimental results validate two critical defensive capabilities. First, we demonstrated that structural deception breaks the fundamental assumption of DPA. By coercing the attacker into selecting an incorrect power model, the correlation peaks required for key recovery are suppressed, yielding a DPA Resilience Score of $\approx 0.45$ (near the theoretical ideal of random guessing). Second, we showed that GNN classifiers are effectively poisoned by the deceptive topology, misclassifying functional logic with high confidence.

Comparison of the proposed methodologies reveals a distinct trade-off: while IP Camouflage (AIG) offers superior area efficiency and maximum structural obfuscation, Graph Matching (GM) provides a formally equivalent, standard-cell compliant alternative suitable for rigorous design flows. Ultimately, these techniques establish a new frontier in hardware security, where protection is achieved not just by hiding the logic, but by actively lying about it.

\section*{Acknowledgment}
This work was supported by the US Army Research Office (ARO) under award \#W911NF-19-1-0102.

\bibliographystyle{IEEEtran}
\bibliography{references}

@inproceedings{torrance2011state,
  title={The state-of-the-art in semiconductor reverse engineering},
  author={Torrance, Randy and James, Dick},
  booktitle={Proceedings of the 48th Design Automation Conference},
  pages={333--338},
  year={2011}
}

@inproceedings{telgarsky2016benefits,
  title={Benefits of depth in neural networks},
  author={Telgarsky, Matus},
  booktitle={Conference on learning theory},
  pages={1517--1539},
  year={2016},
  organization={PMLR}
}

@article{belcak2022neural,
  title={Neural combinatorial logic circuit synthesis from input-output examples},
  author={Belcak, Peter and Wattenhofer, Roger},
  journal={arXiv preprint arXiv:2210.16606},
  year={2022}
}

@article{wang2024towards,
  title={Towards next-generation logic synthesis: A scalable neural circuit generation framework},
  author={Wang, Zhihai and Wang, Jie and Yang, Qingyue and Bai, Yinqi and Li, Xing and Chen, Lei and Hao, Jianye and Yuan, Mingxuan and Li, Bin and Zhang, Yongdong and others},
  journal={Advances in Neural Information Processing Systems},
  volume={37},
  pages={99202--99231},
  year={2024}
}

@inproceedings{erbagci2016secure,
  title={A secure camouflaged threshold voltage defined logic family},
  author={Erbagci, Burak and Erbagci, Cagri and Akkaya, Nail Etkin Can and Mai, Ken},
  booktitle={2016 IEEE International symposium on hardware oriented security and trust (HOST)},
  pages={229--235},
  year={2016},
  organization={IEEE}
}

@inproceedings{prouff2005dpa,
  title={DPA attacks and S-boxes},
  author={Prouff, Emmanuel},
  booktitle={International Workshop on Fast Software Encryption},
  pages={424--441},
  year={2005},
  organization={Springer}
}

@article{wang2024fgnn2,
  title={Fgnn2: A powerful pre-training framework for learning the logic functionality of circuits},
  author={Wang, Ziyi and Bai, Chen and He, Zhuolun and Zhang, Guangliang and Xu, Qiang and Ho, Tsung-Yi and Huang, Yu and Yu, Bei},
  journal={IEEE Transactions on Computer-Aided Design of Integrated Circuits and Systems},
  year={2024},
  publisher={IEEE}
}

@inproceedings{kocher1999differential,
  title={Differential power analysis},
  author={Kocher, Paul and Jaffe, Joshua and Jun, Benjamin},
  booktitle={Annual international cryptology conference},
  pages={388--397},
  year={1999},
  organization={Springer}
}

@article{alrahis2021gnn,
  title={GNN-RE: Graph neural networks for reverse engineering of gate-level netlists},
  author={Alrahis, Lilas and Sengupta, Abhrajit and Knechtel, Johann and Patnaik, Satwik and Saleh, Hani and Mohammad, Baker and Al-Qutayri, Mahmoud and Sinanoglu, Ozgur},
  journal={IEEE Transactions on Computer-Aided Design of Integrated Circuits and Systems},
  volume={41},
  number={8},
  pages={2435--2448},
  year={2021},
  publisher={IEEE}
}

@article{subramanyan2013reverse,
  title={Reverse engineering digital circuits using structural and functional analyses},
  author={Subramanyan, Pramod and Tsiskaridze, Nestan and Li, Wenchao and Gasc{\'o}n, Adria and Tan, Wei Yang and Tiwari, Ashish and Shankar, Natarajan and Seshia, Sanjit A and Malik, Sharad},
  journal={IEEE Transactions on Emerging Topics in Computing},
  volume={2},
  number={1},
  pages={63--80},
  year={2013},
  publisher={IEEE}
}

@inproceedings{meade2016netlist,
  title={Netlist reverse engineering for high-level functionality reconstruction},
  author={Meade, Travis and Zhang, Shaojie and Jin, Yier},
  booktitle={2016 21st Asia and South Pacific Design Automation Conference (ASP-DAC)},
  pages={655--660},
  year={2016},
  organization={IEEE}
}

@article{albartus2020dana,
  title={DANA universal dataflow analysis for gate-level netlist reverse engineering},
  author={Albartus, Nils and Hoffmann, Max and Temme, Sebastian and Azriel, Leonid and Paar, Christof},
  journal={IACR Transactions on Cryptographic Hardware and Embedded Systems},
  pages={309--336},
  year={2020}
}

@inproceedings{principe2017steps,
  title={Steps toward automated deprocessing of integrated circuits},
  author={Principe, Edward L and Asadizanjani, Navid and Forte, Domenic and Tehranipoor, Mark and Chivas, Robert and DiBattista, Michael and Silverman, Scott and Marsh, Mike and Piche, Nicolas and Mastovich, John},
  booktitle={International Symposium for Testing and Failure Analysis},
  volume={81504},
  pages={285--298},
  year={2017},
  organization={ASM International}
}

@article{quadir2016survey,
  title={A survey on chip to system reverse engineering},
  author={Quadir, Shahed E and Chen, Junlin and Forte, Domenic and Asadizanjani, Navid and Shahbazmohamadi, Sina and Wang, Lei and Chandy, John and Tehranipoor, Mark},
  journal={ACM journal on emerging technologies in computing systems (JETC)},
  volume={13},
  number={1},
  pages={1--34},
  year={2016},
  publisher={ACM New York, NY, USA}
}

@article{yu2017incremental,
  title={Incremental SAT-based reverse engineering of camouflaged logic circuits},
  author={Yu, Cunxi and Zhang, Xiangyu and Liu, Duo and Ciesielski, Maciej and Holcomb, Daniel},
  journal={IEEE Transactions on Computer-Aided Design of Integrated Circuits and Systems},
  volume={36},
  number={10},
  pages={1647--1659},
  year={2017},
  publisher={IEEE}
}

@article{RE2016,
title={A survey on chip to system reverse engineering},
  author={Quadir, Shahed E and Chen, Junlin and Forte, Domenic and Asadizanjani, Navid and Shahbazmohamadi, Sina and Wang, Lei and Chandy, John and Tehranipoor, Mark},
  journal={ACM journal on emerging technologies in computing systems (JETC)},
  volume={13},
  number={1},
  pages={1--34},
  year={2016},
  publisher={ACM New York, NY, USA}
}

@misc{arancaGlobalResearch,
	author = {},
	title = {{G}lobal {R}esearch and {A}nalytics {F}irm | {I}nvestment, {B}usiness, {I}ntellectual {P}roperty {R}esearch and {B}usiness {V}aluation {S}ervices | {A}ranca --- aranca.com},
	howpublished = {\url{https://www.aranca.com/knowledge-library/articles/investment-research/ip-theft-china-and-beyond}},
	year = {},
	note = {[Accessed 06-03-2025]},
}

@inproceedings{rajendran2013security,
  title={Security analysis of integrated circuit camouflaging},
  author={Rajendran, Jeyavijayan and Sam, Michael and Sinanoglu, Ozgur and Karri, Ramesh},
  booktitle={Proceedings of the 2013 ACM SIGSAC conference on Computer \& communications security},
  pages={709--720},
  year={2013}
}

@inproceedings{akkaya2018secure,
  title={A secure camouflaged logic family using post-manufacturing programming with a 3.6 ghz adder prototype in 65nm cmos at 1v nominal v dd},
  author={Akkaya, Nail Etkin Can and Erbagci, Burak and Mai, Ken},
  booktitle={2018 IEEE International Solid-State Circuits Conference-(ISSCC)},
  pages={128--130},
  year={2018},
  organization={IEEE}
}

@article{shakya2019covert,
  title={Covert gates: Protecting integrated circuits with undetectable camouflaging},
  author={Shakya, Bicky and Shen, Haoting and Tehranipoor, Mark and Forte, Domenic},
  journal={IACR transactions on cryptographic hardware and embedded systems},
  pages={86--118},
  year={2019}
}

@article{fan2025designing,
  title={Designing with Deception: ML-and Covert Gate-Enhanced Camouflaging to Thwart IC Reverse Engineering},
  author={Fan, Junling and Koblah, David and Forte, Domenic},
  journal={arXiv preprint arXiv:2508.08462},
  year={2025}
}

@article{fan2025scalable,
  title={Scalable IP Mimicry: End-to-End Deceptive IP Blending to Overcome Rectification and Scale Limitations of IP Camouflage},
  author={Fan, Junling and Rushevich, George and Rusconi, Giorgio and Zhu, Mengdi and Dizon-Paradis, Reiner and Forte, Domenic},
  journal={arXiv preprint arXiv:2512.12061},
  year={2025}
}

\end{document}